\documentclass[useAMS,usenatbib]{mn2e}

\usepackage{graphicx}

\newcommand       \be           {\begin{equation}}
\newcommand       \ee           {\end{equation}}
\newcommand       \bea          {\begin{eqnarray}}
\newcommand       \eea          {\end{eqnarray}}
\newcommand       \apj          {ApJ}
\newcommand       \apjl         {ApJL}
\newcommand       \aap          {A\&A}

\newcommand       \mnras        {MNRAS}
\def\simlt{\mathrel{\hbox{\rlap{\hbox{\lower4pt\hbox{$\sim$}}}\hbox{$<$}}}}
\def\simgt{\mathrel{\hbox{\rlap{\hbox{\lower4pt\hbox{$\sim$}}}\hbox{$>$}}}}
\def\lesssim{\mathrel{\hbox{\rlap{\hbox{\lower4pt\hbox{$\sim$}}}\hbox{$<$}}}}

\def\gtrsim{\mathrel{\hbox{\rlap{\hbox{\lower4pt\hbox{$\sim$}}}\hbox{$>$}}}}

\title[Neutron-Rich Freeze-Out in Accretion Disks]{Neutron-rich freeze-out in viscously spreading accretion disks formed from compact object mergers} \author[B.~D. Metzger, A.~L. Piro, E. Quataert]{B.~D. Metzger\thanks{E-mail:
bmetzger@astro.berkeley.edu}, A.~L. Piro, and E. Quataert \\
Astronomy Department and Theoretical Astrophysics Center,
University of California, Berkeley, 601 Campbell Hall, Berkeley CA,
94720\\}
\begin{document}
\date{Accepted . Received ; in original form }
\pagerange{\pageref{firstpage}--\pageref{lastpage}} \pubyear{????}
\maketitle
\label{firstpage}

\begin{abstract}
Accretion disks with masses $\sim 10^{-3}-0.1M_{\sun}$ are believed to form during the merger of a neutron star (NS) with another NS and the merger of a NS with a black hole (BH).  Soon after their formation, such hyper-accreting disks cool efficiently by neutrino emission and their composition is driven neutron-rich by pair captures under degenerate conditions.  However, as the disk viscously spreads and its temperature drops, neutrino cooling is no longer able to offset viscous heating and the disk becomes advective.  Analytic arguments and numerical simulations suggest that once this occurs, powerful winds likely drive away most of the disk's remaining mass.  We calculate the thermal evolution and nuclear composition of viscously spreading accretion disks formed from compact object mergers using one-dimensional height-integrated simulations.  We show that freeze-out from weak equilibrium necessarily accompanies the disk's late-time transition to an advective state.  As a result, hyper-accreting disks generically freeze out neutron-rich (with electron fraction $Y_{e} \sim 0.2-0.4$), and their late-time outflows robustly synthesize rare neutron-rich isotopes.  Using the measured abundances of these isotopes in our solar system, we constrain the compact object merger rate in the Milky Way to be $\lesssim 10^{-5}(M_{d,0}/0.1M_{\sun})^{-1}$ yr$^{-1}$, where $M_{d,0}$ is the average initial mass of the accretion disk.  Thus, either the NS-NS merger rate is at the low end of current estimates or the average disk mass produced during a typical merger is $\ll 0.1 M_{\sun}$.  Based on the results of current general relativistic merger simulations, the latter constraint suggests that prompt collapse to a BH is a more common outcome of NS-NS mergers than the formation of a transient hyper-massive NS.  We also show that if most short duration gamma-ray bursts (GRBs) are produced by compact object mergers, their beaming fraction must exceed $f_{b} \approx 0.13(M_{d,0}/0.1 M_{\sun})$, corresponding to a jet half-opening angle $\gtrsim 30^{\circ}(M_{d,0}/0.1M_{\sun})^{1/2}$.  This is consistent with other evidence that short duration GRB outflows are less collimated than those produced in long duration GRBs.
\end{abstract}

\begin{keywords}
{nuclear reactions, nucleosynthesis, abundances -- accretion disks ---
	black hole physics ---
	gamma rays: bursts ---
	neutrinos}
\end{keywords}

\section{Introduction}
\voffset=-2cm
\vspace{0.2 cm}
\label{sec:int}
\label{sec:intro}

Massive, compact accretion disks are thought to form in a number of astrophysical events, including the merger of a black hole and a neutron star (BH-NS; Lattimer $\&$ Schramm 1974, 1976), the merger of a double NS binary (NS-NS; Eichler et al.~1989), the collapse of a rapidly-rotating stellar core (Woosley 1993), the accretion-induced collapse (AIC) of a white dwarf (WD) to a NS (Dessart et al.~2006, 2007), and the AIC of a NS to a BH (Vietri $\&$ Stella 1998, 1999).  These disks are termed ``hyper-accreting'' due to their large accretion rates of up to several solar masses per second.  Rapid accretion onto a BH following the collapse of a massive star is a popular model for the central engine of long-duration gamma-ray bursts (GRBs) (MacFadyen $\&$ Woosley 1999) due to their association with star forming regions and Type Ibc supernovae (SNe) (Woosley $\&$ Bloom 2006).  Short-duration GRBs, on the other hand, originate from a more evolved progenitor population (e.g., Berger et al.~2005) and may instead result from compact object (CO) mergers (e.g., Lee $\&$ Ramirez-Ruiz 2007), or the AIC of a WD (Metzger et al.~2008a) or a NS (MacFadyen et al.~2005).

Hyper-accreting disks cool via neutrino emission when the accretion rate is sufficiently high ($\dot{M} \gtrsim 0.01-0.1 M_{\sun}$ s$^{-1}$; Popham et al.~1999).  This allows the disk midplane to become dense and electron degenerate (Chen $\&$ Beloborodov 2007).  Because positron captures ($e^{+}+n\rightarrow p+\bar{\nu}_{e}$) are suppressed relative to electron captures  ($e^{-}+p\rightarrow n+\nu_{e}$) under degenerate conditions, neutrino-cooled disks are driven neutron-rich (Pruet et al.~2003; Beloborodov 2003), i.e., to an electron fraction $Y_{e} \equiv n_{\rm p}/(n_{\rm p} + n_{\rm n}) < 0.5$, where $n_{\rm p}$ and $n_{\rm n}$ are the proton and neutron density, respectively.  

If a portion of this neutron-rich material escapes the disk, several observable consequences may result (e.g., Derishev et al.~1999).  In particular, although highly relativistic outflows are required to produce GRBs, slower outflows from larger radii in the disk are probably at least as common (e.g., Pruet et al.~2004).  These dense outflows may synthesize heavy, neutron-rich isotopes as they expand away from the midplane and cool.  Such outflows are difficult to detect directly because neutron-rich isotopes have short half-lives ($\sim$ seconds; Freiburghaus et al.~1999) and are thus unlikely to power a bright SN-like transient via their radioactive decay (although a dimmer, shorter-lived transient may be produced; Li $\&$ Paczynski 1998; Kulkarni 2005; Metzger, Piro, $\&$ Quataert 2008, hereafter MPQ08).  

Neutron-rich outflows can also be probed indirectly via their effect on the chemical evolution of the Galaxy.  For instance, moderately neutron-rich outflows with $Y_{e} \approx 0.35-0.4$ produce isotopes (e.g., $^{79}$Br and $^{78}$Se) that are rare in our solar system (Hartmann et al.~1985; Woosley $\&$ Hoffman 1992).  Hence, the measured abundances of these isotopes place stringent constraints on the rate at which low-$Y_{e}$ material is ejected into the interstellar medium (ISM).  In turn, if the amount of neutron-rich material ejected in a given event can be estimated, the true rate and beaming fraction (or jet opening angle) could in principle be constrained in models that associate these events with GRBs (e.g., Woosley $\&$ Baron 1992; Fryer et al.~1999).  Constraints on the rate of CO mergers are particularly interesting because they are the primary target for km-scale gravitational wave detectors such as LIGO and VIRGO, and rates derived from known merging systems (Kalogera et al.~2004; Kim et al.~2006) and population synthesis models (e.g., Belczynski et al.~2006) remain highly uncertain.

\subsection{Summary of Previous Work}

Most previous efforts to determine the composition of outflows from hyper-accreting disks have focused on winds driven by neutrino irradiation from the inner disk (Levinson 2006; Surman et al.~2006, 2008; Barzilay $\&$ Levinson 2007; Metzger et al.~2008b), in analogy to those from proto-neutron stars following successful core-collapse SNe (e.g., Burrows, Hayes, $\&$ Fryxell 1995).  However, the electron fraction at the base of neutrino-driven outflows is typically not preserved as matter accelerates to large radii because neutrino absorptions ($p+\bar{\nu}_{e}\rightarrow n+e^{+}$ and $n+\nu_{e}\rightarrow p+e^{-}$) drive $Y_{e}$ to the equilibrium value 
\be Y_{e}^{\nu} \approx [1 + (L_{\bar{\nu}_{e}}\langle \epsilon_{\bar{\nu}_{e}}\rangle/L_{\nu_{e}}\langle\epsilon_{\nu_{e}}\rangle)]^{-1}
\label{eq:yenu}
\ee given by the properties of the neutrino radiation field, where $L_{\nu_{e}}/L_{\bar{\nu}_{e}}$ and $\langle\epsilon_{\nu_{e}}\rangle$/$\langle\epsilon_{\bar{\nu}_{e}}\rangle$ are the mean $\nu_{e}$/$\bar{\nu}_{e}$ luminosities and energies, respectively, from the central NS or accretion disk (Qian et al.~1993).  When the accretion rate is very high ($\dot{M} \gtrsim M_{\sun}$ s$^{-1}$) and the inner disk is optically thick to neutrinos, $Y_{e}^{\nu}$ can itself be $\ll 0.5$ because the $\bar{\nu}_{e}$'s originate from regions of higher temperature than the $\nu_{e}$'s (e.g., Surman et al.~2006).  However, a disk with a fixed initial mass (such as is produced by a CO merger) spends very little time (if any) accreting at such a high rate; most neutrino-driven mass loss from a viscously-spreading disk occurs at later times, when the entire disk is optically thin and $Y_{e}^{\nu} \sim 0.5$ (see Fig.~9 of MPQ08).  Thus, while early-time neutrino-driven outflows may produce modest amounts of $r$-process ejecta, their total yield in the context of CO mergers is probably insufficient to contribute appreciably to the Galactic abundance of neutron-rich isotopes (e.g., Surman et al.~2008).  However, as we now discuss, later stages in the disk's evolution, when neutrino irradiation is relatively unimportant, are likely to eject even larger quantities of neutron-rich material.

In a recent paper (MPQ08), we used a one-zone (or ``ring'') model to study the evolution of hyper-accreting disks.  Guided by analytic work (Blandford $\&$ Begelman 1999) and numerical simulations (e.g., Stone $\&$ Pringle 2001; Hawley, Balbus, $\&$ Stone 2001; Hawley $\&$ Balbus 2002) which show that radiatively-inefficient accretion drives powerful outflows, we argued that the majority of the disk becomes unbound soon following the disk's transition to an advective state at late times.  Since ejection occurs on approximately the dynamical timescale (which is faster than weak interactions at this stage), these ``viscously-driven'' outflows maintain the electron fraction of the disk midplane, unlike the neutrino-driven outflows at earlier times.  Thus, in order to determine what isotopes are ultimately synthesized, the electron fraction in the disk must be known at late times when the disk becomes advective.  

\subsection{Outline of this Paper}

In this paper we determine the composition of late-time outflows from hyper-accreting disks by calculating the disk's midplane composition as the disk viscously spreads and falls out of weak equilibrium.  We concentrate in particular on BH accretion following a CO merger\footnote{Although we center our discussion on accretion following CO mergers, most of our conclusions would apply equally to the AIC of a NS to a BH, which may produce a disk with similar initial properties (Shibata 2003; Shapiro 2004).} (either NS-NS or NS-BH models), where a fixed initial disk mass is a reasonable approximation.  In $\S\ref{sec:onezonemodel}$ we use a one-zone model similar to that in MPQ08 to calculate $Y_{e}(t)$ for a wide variety of plausible initial disk masses and sizes.  Although the one-zone model in $\S\ref{sec:onezonemodel}$ captures the basic evolution of hyper-accreting disks, the subsequent nucleosynthesis depends sensitively on the freeze-out electron fraction $Y_{e}^{f}$, and so a more detailed calculation is warranted.  Furthermore, a one-zone model cannot, by construction, address the possibility that different annuli in the disk may freeze out with different values of $Y_{e}^{f}$.  Therefore in $\S\ref{sec:onedmodel}$ we present one-dimensional (1D), height-integrated calculations of the evolution of CO merger disks and their composition, and use them to determine the electron fraction mass distribution $M(Y_{e}^{f})$ at late times.  We find that under most conditions the majority of the mass freezes out neutron-rich with $Y_{e}^{f} \sim 0.2-0.4$.  In $\S\ref{sec:discussion}$ we summarize our results and use them to constrain the CO merger rate and the beaming fraction of short GRBs.

\section{One Zone Model}
\label{sec:onezonemodel}

In this section we present calculations of the electron fraction at freeze-out $Y_{e}^{f}$ using a one-zone ``ring'' model similar to that presented in MPQ08.  The simplicity of this model allows us to efficiently explore a wide parameter space of initial disk mass $M_{d,0}$ and radius $r_{d,0}$, and it also provides a useful point of comparison for our 1D calculations in $\S\ref{sec:onedmodel}$.  The initial conditions and relevant equations are presented in $\S\ref{sec:onezoneequations}$.  In $\S\ref{sec:onezoneresults}$ we present our results and show, using a simple analytic argument, that a moderately neutron-rich freeze-out (i.e., $Y_{e}^{f} \lesssim 0.5$) is generically expected, relatively independent of the details of how the disk viscously spreads.  

\subsection{Equations and Initial Conditions}
\label{sec:onezoneequations}

Since the accretion disks produced during CO mergers are created from tidally disrupted NS material, we take the initial electron fraction to be $Y_{e}^{0} = 0.1$, which is characteristic of the inner neutron star crust (e.g., Haensel $\&$ Zdunik 1990a,b; Pethick $\&$ Ravenhall 1995).  In most of our calculations we take the BH mass to be $M_{\rm BH} = 3 M_{\sun}$.  The mass and radial profile of the remnant disks formed from CO mergers are uncertain theoretically because they depend on the unknown supranuclear-density equation of state and general relativistic effects, which are now being explored in merger simulations (e.g., Shibata $\&$ Taniguchi 2006).  Disk masses $M_{d} \sim 10^{-3}-0.3 M_{\sun}$ with characteristic sizes $\sim 10^{6}-3\times 10^{7}$ cm appear typical (e.g., Rasio et al.~2005; Oechslin $\&$ Janka 2006; Shibata $\&$ Taniguchi 2006).
     
The equations and assumptions employed in our one-zone model closely follow those in MPQ08.  To provide a brief summary, our calculation follows the thermal and viscous evolution of the disk radius $r_{d}(t)$ (the ``ring'') that contains the majority of the mass, controls the accretion rate $\dot{M}$ onto the central object, and moves outwards as the disk accretes in order to conserve the total angular momentum $J \propto M_{d}r_{d}^{1/2}$.  Our energy equation includes viscous heating and the dominant neutrino opacities.  The model is calibrated so as to reproduce the exact $\delta$-function solution to the viscous diffusion equation (eq.~[\ref{eq:sigma_evo}]) at late times.  Although we refer the reader to MPQ08 for details, we include here a brief discussion of changes and additions we have made and highlight the equations most important for this work.  

The electron fraction is evolved using 
\bea
\frac{dY_e}{dt} = (\lambda_{e^+n}+\lambda_{\nu_{e}n})\left[1-Y_{e}-\left(\frac{1-X_{\rm f}}{2}\right)\right] \nonumber \nonumber
\eea
\be
-(\lambda_{e^{-}p}+\lambda_{\bar{\nu}_{e}p})\left[Y_{e}-\left(\frac{1-X_{\rm f}}{2}\right)\right],
\label{eq:yeevo}
\ee
where $d/dt$ is a Lagrangian derivative, $\lambda_{e^{-}p}/\lambda_{e^{+}n}$ are the pair capture rates (see Beloborodov 2003, eqs.~[6], [7]), $X_{\rm f}$ is the free nucleon mass fraction in nuclear statistical equilibrium (NSE) (from, e.g., Woosley $\&$ Baron 1992), and 
\be
\lambda_{\nu N} = \frac{L_{\nu}\sigma_{\nu N}X_{N}}{4\pi r_{d}^{2} \langle\epsilon_{\nu}\rangle(1+\tau_{\nu N})} \approx 500\frac{X_{N}L_{52}\langle\epsilon_{10}\rangle}{r_{6}^{2}(1+\tau_{\nu N})}{\rm\, s^{-1}}
\label{eq:nucaptrate}
\ee
are the neutrino capture rates (e.g., Qian $\&$ Woosley 1996) due to irradiation from the inner accretion disk (see the discussion below).  In equation (\ref{eq:nucaptrate}), $\nu N$ stands for either $\nu_{e}n$ or $\bar{\nu}_{e}p$, $X_{N}$ is the corresponding proton or neutron mass fraction, $L_{\nu} \equiv  L_{52}10^{52}$ ergs s$^{-1}$ is the neutrino luminosity, $\langle\epsilon_{\nu}\rangle \equiv 10 \langle\epsilon_{10}\rangle$ MeV is the mean neutrino energy, and $\sigma_{\nu N} \simeq 9\times 10^{-44}\langle\epsilon_{\nu}^{2}\rangle$ MeV$^{-2}$cm$^{2}$ is the neutrino capture cross section.  The factor (1 + $\tau_{\nu N}$) accounts for the possibility that the absorbing annulus may become optically thick, where $\tau_{\nu N} \equiv \sigma_{\nu N}X_{N}\rho r_{d}/2m_{p}$ is the neutrino optical depth and $\rho$ is the density; our calculation is fairly insensitive to this prescription for when $\tau_{\nu} > 1$, however, because $\tau_{\nu}$ is generally $\ll 1$ at freeze-out.  We neglect the proton-neutron mass difference $\Delta \approx 1.3$ MeV and the effects of electron degeneracy in calculating $\sigma_{\nu N}$ because the energies of the neutrinos that dominate the heating ($\langle\epsilon_{\nu}\rangle \sim 10$ MeV) greatly exceed $\Delta$ and the electron Fermi energy, respectively.  We assume relativistic $e^{-}/e^{+}$ pairs when calculating the electron chemical potential; we have verified that this is a good approximation by checking that including the effects of arbitrary $e^{-}/e^{+}$ energies has no significant effect on the value of $Y_{e}^{f}$.  NSE is a generally a good assumption in calculating $X_{f}$ because the entropy in the disk is sufficiently low that $\alpha-$particles form while the midplane temperature is still high ($T \gtrsim 0.5$ MeV).

The dominant process that sets $Y_{e}$ during the thin disk phase is pair capture on free nuclei (e.g., Pruet et al.~2003).  However, neutrino absorptions become important near freeze-out and must be included in a detailed calculation (see Fig.~\ref{plot:rates_onezone}).  The neutrino absorption rate $\lambda_{\nu N}$ is dominated by the high energy neutrinos that are radiated from the smallest radii in the disk.  We calculate $L_{\nu}$ and $\langle\epsilon_{\nu}\rangle$ by extending our solutions down to the innermost stable orbit $r_{*}$, assuming that $\dot{M}$ is constant with radius interior to $r_{d}$.  The neutrino flux is reasonably approximated as a ``light bulb'' at the origin in equation (\ref{eq:nucaptrate}) because the neutrino capture rates only become comparable to the pair capture rates when the disk has spread to a radius that is much larger than that of the central BH.

Since neutrino captures only affect $Y_{e}$ appreciably at times when the inner disk is optically thin to neutrinos, we assume that $L_{\nu_{e}} \approx L_{\bar{\nu}_{e}}$ and $\langle\epsilon_{\nu_{e}}\rangle \approx \langle\epsilon_{\bar{\nu}_{e}}\rangle$ (see the discussion in $\S$5.3 of MPQ08 for why this is a good approximation).  Thus, the effect of neutrino captures is to drive $Y_{e}$ towards $Y_{e}^{\nu} \approx 0.5$ (see eq.~[\ref{eq:yenu}]).  The effect of $\alpha$-particle formation (i.e., $X_{f} \rightarrow 0$ in eq.~[\ref{eq:yeevo}]) would also be to drive $Y_{e} \rightarrow 0.5$ (the ``alpha-effect''; e.g., Fuller $\&$ Meyer 1995); however, this effect is unimportant in our calculations because weak freeze-out generally precedes $\alpha$-particle formation.

One difference relative to MPQ08 is that here we take the viscous stress to be proportional to just the ion gas pressure $P_{\rm gas}$, as opposed to the total pressure $P_{\rm tot}$, which also includes radiation pressure from photons and $e^{-}/e^{+}$ pairs.  In other words, we take the kinematic viscosity to be 
\be
\nu = \alpha P_{\rm gas}/\rho\Omega,
\label{eq:viscosity}
\ee
where $\alpha$ is a dimensionless constant and $\Omega$ is the Keplerian rotation rate.  At early times in the evolution of the disk (in particular when it is neutrino-cooled), the disk is primarily supported by gas pressure and so using either $\nu \propto P_{\rm tot}$ or $\nu \propto P_{\rm gas}$ gives similar results.  However, as the disk becomes thick at late times and $Y_{e}$ freezes out, the disk becomes dominated by radiation pressure, with $P_{\rm tot}/P_{\rm gas} \sim 2-3$ during freeze-out (see Fig.~\ref{plot:rates_onezone}).  The primary reason that we use $P_{\rm gas}$ instead of $P_{\rm tot}$ in equation (\ref{eq:viscosity}) is to avoid the classic Lightman-Eardley (1974) viscous instability, which we otherwise find develops in our full 1D calculations described in $\S\ref{sec:onedmodel}$.  Because $P_{\rm tot}/P_{\rm gas}$ is modest ($\lesssim 3$) during the times that matter to our results, we choose $\alpha \sim 0.3$ and $0.03$ as fiducial values for our calculations, to roughly bracket the value of $\alpha \sim 0.1$ motivated by a number of astrophysical observations (King et al.~2007).  For typical disk parameters, we find that taking $\nu \propto P_{\rm gas}$ instead of $\nu \propto P_{\rm tot}$ in our one-zone models results in a modest ($\sim 20\%$) increase in the final electron fraction.  This, together with our inclusion of neutrino absorptions in evolving $Y_{e}$, accounts for the fact that the final electron fractions $Y_{e}^{f}$ presented here are somewhat larger than those presented in MPQ08 (their Fig.~9).

As an improvement over MPQ08, our calculation of the cooling rates due to $e^{-}/e^{+}$ captures on free nuclei now includes the effects of arbitrary electron energy and degeneracy.  In addition, we include heating from the absorption of neutrinos that are radiated from the accretion disk at small radii (see eq.~[\ref{eq:nuheating}]).  Finally, as in MPQ08 we apply a ``no torque'' boundary condition at the inner radius $r_{*}= 10^{6}$ cm, which corresponds to the radius of the innermost stable orbit around a rapidly rotating Kerr BH ($a = 0.9$).

\subsection{Results}
\label{sec:onezoneresults}

\begin{figure}
\begin{center}
\resizebox{\hsize}{!}{\includegraphics[ ]{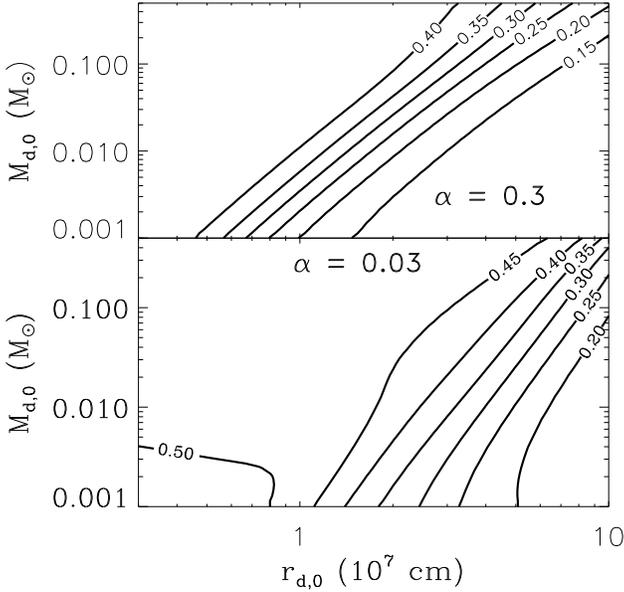}}
\end{center}
\caption{Contours of the final electron fraction $Y_{e}^{f}$ following weak freeze-out in CO merger disks as a function of the initial disk mass $M_{d,0}$ and radius $r_{d,0}$ for two values of the viscosity, $\alpha = 0.3$ ({\it top}) and $\alpha = 0.03$ ({\it bottom}).  The initial electron fraction in all models is taken to be $Y_{e}^{0} = 0.1$.}  
\label{plot:yef_onezone}
\end{figure}

\begin{figure*}
\begin{center}$
\begin{array}{cc}
\includegraphics[width=8.5cm]{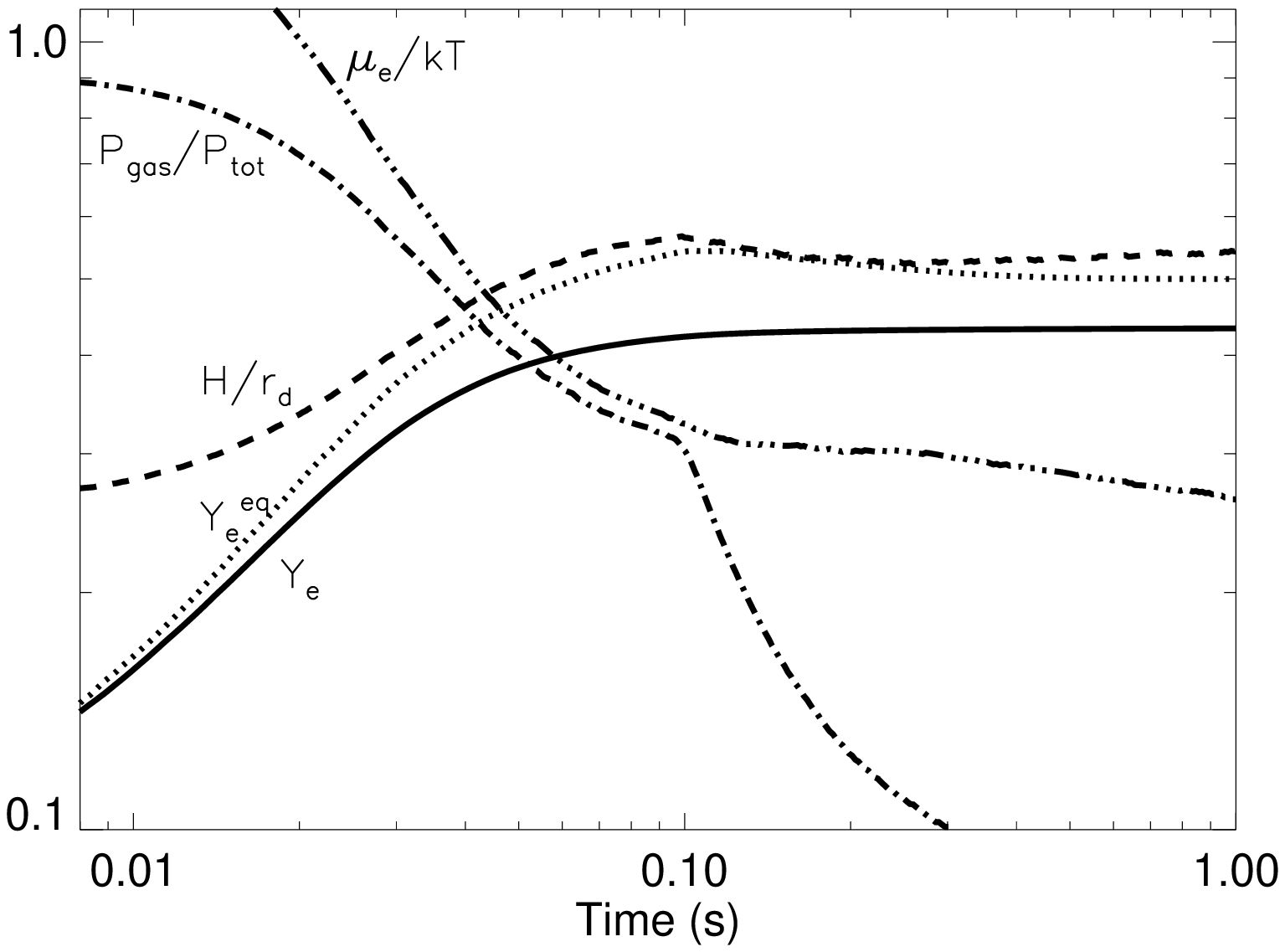} &
\includegraphics[width=8.5cm]{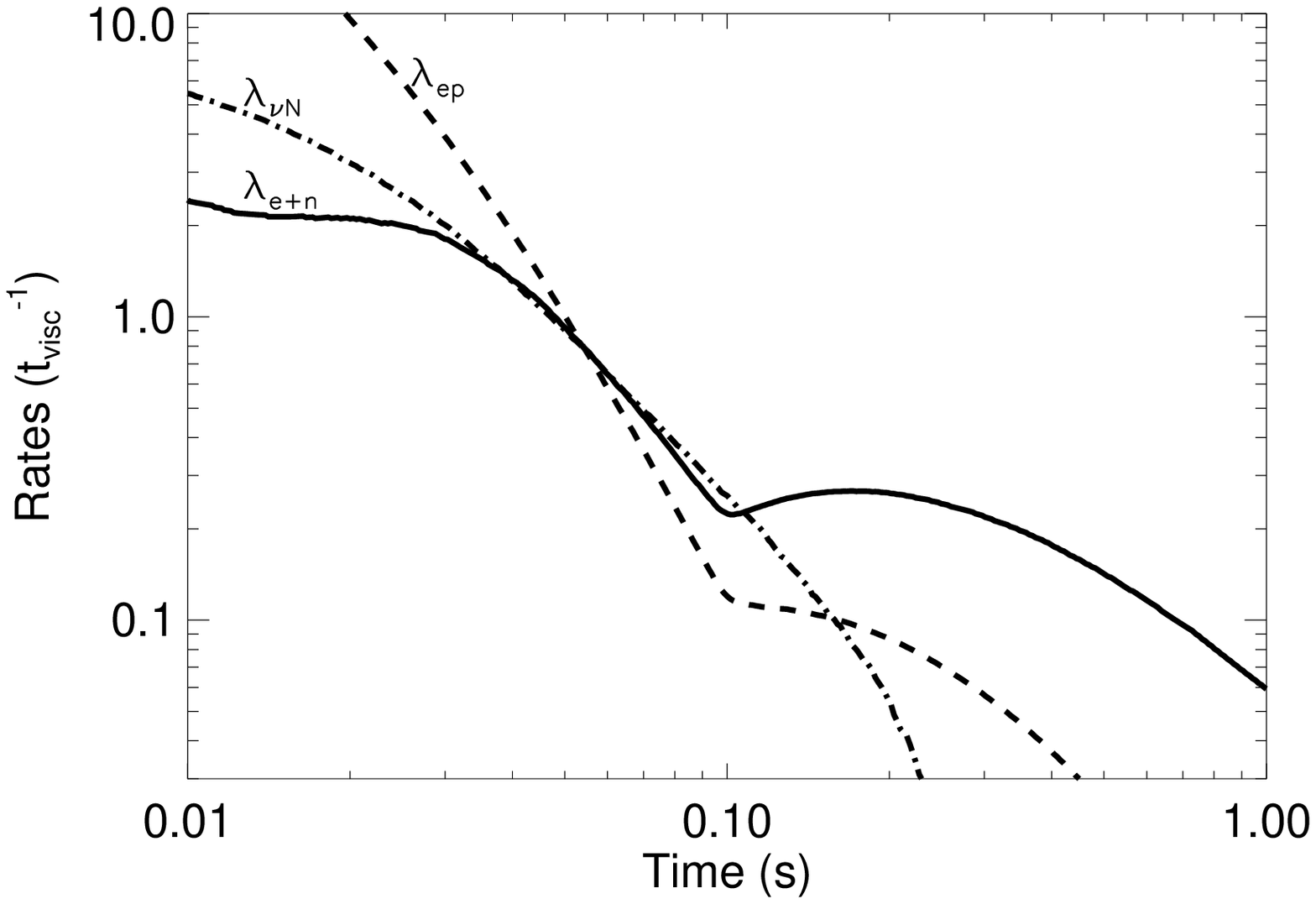}
\end{array}$
\end{center}
\caption{The process of weak freeze-out in our one-zone model of a viscously-spreading, hyper-accreting disk.  The left panel shows the time evolution of the electron fraction $Y_{e}$ ({\it solid line}), disk thickness $H/r_{d}$ ({\it dashed line}), degeneracy parameter $\mu_{e}/kT$ ({\it triple-dot-dashed line}), and the ratio of gas to total pressure $P_{\rm gas}/P_{\rm tot}$ ({\it dot-dashed line}), calculated for $\alpha = 0.3$ and for a disk with initial mass $M_{d,0} = 0.1M_{\sun}$ and radius $r_{d,0} = 3\times 10^{6}$ cm.  Also shown with a dotted line is the equilibrium electron fraction $Y_{e}^{eq}$ obtained by setting the right hand side of equation (\ref{eq:yeevo}) to zero.  The right panel shows the rates of the weak interactions that modify $Y_{e}(t)$, normalized to the viscous timescale $t_{\rm visc}$.  The electron, positron, and neutrino capture rates are denoted by $\lambda_{ep}$, $\lambda_{e^{+}n}$, and $\lambda_{\nu N}$, respectively.  Note that $Y_{e}$ freezes out of equilibrium at the same time that the disk becomes advective ($H/r_{d} \gtrsim 0.5$) and non-degenerate ($\mu_{e} \lesssim kT$).  Since $Y_{e}^{eq}$ rises to $\gtrsim 0.5$ as the disk becomes non-degenerate, the final composition is only moderately neutron-rich, with a final electron fraction $Y_{e}^{f} \simeq 0.43$.}  
\label{plot:rates_onezone}
\end{figure*}
Figure \ref{plot:yef_onezone} shows contour plots of the freeze-out electron fraction $Y_{e}^{f}$ as a function of the initial disk mass $M_{d,0}$ and radius $r_{d,0}$ using two different values of the viscosity, $\alpha = 0.03$ and $0.3$.

The basic features of Figure \ref{plot:yef_onezone} can be understood as follows.  Sufficiently massive, compact disks (upper left corner of Fig.~\ref{plot:yef_onezone}) come into pair capture-equilibrium and obtain an electron fraction that is relatively independent of the initial value $Y_{e}^{0} = 0.1$, while low mass, more extended disks (lower right corner) retain their initial composition.  As discussed below, a disk enters $\beta-$equilibrium if and only if it can cool efficiently via neutrino emission (see eq.~[\ref{eq:foargument}]).  By equating the radius where neutrino cooling balances $\sim 1/2$ of the viscous heating in a self-similar solution for a thick disk, we estimate analytically that disks that satisfy
\be 
M_{d,0} \gtrsim 3\times 10^{-3}\left(\frac{r_{d,0}}{10^{7}{\rm cm}}\right)^{7/3}\left(\frac{\alpha}{0.1}\right)^{2/3}M_{\sun} 
\label{eq:md0}
\ee
are sufficiently compact to become neutrino-cooled and neutron-rich at early stages in their evolution (see also MPQ08 eqs.~[19]-[21]).  Equation (\ref{eq:md0}) does a reasonable job of reproducing the parameter space in Figure \ref{plot:yef_onezone} where $Y_{e}^{f}$ differs significantly from its initial value.  For instance, equation (\ref{eq:md0}) shows that lower values of $\alpha$ allow less massive disks of a fixed size to enter $\beta$-equilibrium; this is consistent with the wider parameter space with $Y_{e}^{f}>Y_{e}^{0}=0.1$ at low $\alpha$.

To illustrate the process of weak freeze-out explicitly, Figure \ref{plot:rates_onezone} ({\it left panel}) shows the evolution of the electron fraction $Y_{e}(t)$ for a disk with $M_{d,0} = 0.1 M_{\sun}$ and $r_{d,0} = 3\times 10^{6}$ cm.  We also show the equilibrium electron fraction $Y_{e}^{eq}$ (obtained by setting the right hand side of eq.~[\ref{eq:yeevo}] to zero), the ratio of the disk scaleheight to the radius $H/r_{d}$, the ratio of the gas pressure to the total pressure $P_{\rm gas}/P_{\rm tot}$, and the ratio of the electron chemical potential to the disk temperature $\mu_{e}/kT$ (i.e., the degeneracy parameter).  Also shown for comparison in the right panel of Figure \ref{plot:rates_onezone} are the weak interaction rates from equation (\ref{eq:yeevo}), normalized to the viscous timescale $t_{\rm visc} \equiv \alpha^{-1}\Omega^{-1}(H/r_{d})^{-2}$.

Figure \ref{plot:rates_onezone} shows that at early times, when the disk is neutrino-cooled and $H/r_{d} \sim 1/3$, pair captures are rapid compared to the viscous timescale.  Thus, the disk enters $\beta-$equilibrium with an electron fraction $Y_{e} \approx Y_{e}^{eq}$ that is $\ll 0.5$ because positron captures are suppressed relative to electron captures under the degenerate conditions ($\mu_{e} \gg kT$) in the midplane.  At later times ($t \sim 0.1$ s) the disk thickens to $H/r_{d} \sim 0.5$ and becomes non-degenerate ($\mu_{e} \ll kT$) and radiation pressure-dominated.  At this point $Y_{e}^{\rm eq}$ rises to $\gtrsim 0.5$ because positron captures under non-degenerate conditions now become energetically favored over electron captures due to the proton-neutron mass difference.  Note, however, that as the disk thickens weak interactions become slow compared to the viscous timescale and $Y_{e}$ freezes out at the value $Y_{e}^{f} \approx 0.43$.  Although (non-local) neutrino absorptions are unimportant relative to electron captures at early times, their rate $\lambda_{\nu N}$ is comparable to the pair capture rates near freeze-out and must be included to accurately determine $Y_{e}^{f}$. 

Figure \ref{plot:yef_onezone} shows that moderately neutron-rich freeze-out ($Y_{e}^{f} \sim 0.3-0.5$) is a common feature of accretion disks formed from CO mergers for a wide range of initial conditions.  However, contrasting the disk's neutron-rich equilibrium state ($Y_{e} \simeq Y_{e}^{eq} \sim 0.1$) at early times with its final, non-degenerate state that actually favors a {\it proton-rich} composition in equilibrium (i.e., $Y_{e}^{eq} > 0.5$), it is {\it a priori} unclear why $Y_{e}$ freezes out with a value $Y_{e}^{f} \sim 0.3-0.5$ between these two extremes.  

To understand why this is the case, note that an approximate condition for weak freeze-out is that the pair capture timescale becomes longer than the viscous time $t_{\rm visc}$.  Since the disk becomes non-degenerate around freeze-out, electron and positron captures occur at a similar rate (e.g., Qian $\&$ Woosley 1996)
\be
	\lambda_{eN} \approx 0.45T_{\rm MeV}^{5}{\rm\,s^{-1}},
\label{eq:captrate}
\ee
where $T \equiv T_{\rm MeV}$MeV and $eN$ represents either $e^{-}p$ or $e^{+}n$.  The associated URCA cooling rate per nucleon is given by
\be 	
	\dot{Q}_{eN} = \langle\epsilon_{\nu}\rangle\lambda_{eN} \approx 2.3T_{\rm MeV}^{6}{\rm\,MeV\,s^{-1}}, 
\label{eq:urcacool}
\ee
where $\langle\epsilon_{\nu}\rangle \approx 5.04kT$ is the mean neutrino energy released per capture.  The viscous heating rate per nucleon at $r \gg r_{*}$ is
\be
	\dot{Q}_{\rm visc} = \frac{9}{4}\nu\Omega^{2}m_{\rm N} \approx \frac{9}{4}\alpha m_{\rm N}r^{2}\Omega^{3}\left(\frac{H}{r}\right)^{2},
\label{eq:vischeat}
\ee
where $m_{\rm N}$ is the nucleon mass.  If we assume that the disk is supported primarily by gas pressure\footnote{As discussed in $\S\ref{sec:onezoneequations}$ and shown in Figure \ref{plot:rates_onezone}, gas and radiation pressure are comparable at freeze-out; the assumption that gas pressure dominates is, however, sufficient for the purposes of a simple estimate.} then the midplane temperature is $kT \approx m_{\rm N}r^{2}\Omega^{2}(H/r)^{2}$.  Combining this with equations $(\ref{eq:captrate})-(\ref{eq:vischeat})$ we find that 
\bea
 \lambda_{eN}t_{\rm visc} = \nonumber
\eea
\be
 \frac{9}{4}\left(\frac{kT}{\langle\epsilon_{\nu}\rangle}\right)\left(\frac{H}{r}\right)^{-2}\left(\frac{\dot{Q}_{eN}}{\dot{Q}_{\rm visc}}\right) \simeq 1.8\left(\frac{\dot{Q}_{eN}}{\dot{Q}_{\rm visc}}\right)\left(\frac{H/r}{0.5}\right)^{-2}.
\label{eq:foargument}
\ee
Equation (\ref{eq:foargument}) shows that once neutrino cooling no longer offsets viscous heating (i.e., $\dot{Q}_{\rm visc} \gtrsim \dot{Q}_{eN}$) and the disk thickens to $H/r \sim 1$, weak freeze-out (i.e., $\lambda_{eN}t_{\rm visc} \lesssim 1$) necessarily results.  Physically, this occurs because the disk is cooled by $e^{-}/e^{+}$ captures, the very same processes that largely control the evolution of $Y_{e}$ (see Fig.~\ref{plot:rates_onezone}).  This conclusion is fairly robust because it applies at any radius and is independent of the value of $\alpha$.  Although the absorption of neutrinos from the central disk threatens to complicate this simple argument, the right panel of Figure \ref{plot:rates_onezone} shows that the effect of absorptions is at most comparable to that of pair captures prior to freeze-out.

\section{Height-Integrated Model}
\label{sec:onedmodel}

The results of $\S\ref{sec:onezonemodel}$ suggest that neutron-rich freeze-out is a common feature of viscously-spreading, hyper-accreting disks.  However, the one-zone model makes a number of simplifying assumption that may affect the final electron fraction $Y_{e}^{f}$.  In this section we present a 1D (height-integrated) model of a viscously-spreading disk that more precisely determines $Y_{e}^{f}$ and its distribution with mass.  In $\S\ref{sec:setup}$ we describe the initial conditions and relevant equations.  Then we present our results in $\S\ref{sec:COMresults}$. 

\subsection{Equations and Initial Conditions}
\label{sec:setup}

Our 1D calculation evolves the surface density $\Sigma$, midplane temperature $T$, and electron fraction $Y_{e}$ as a function of radius $r$ and time $t$ using the 2N-RK3 (6th order space, 3rd order time) scheme described in Brandenburg (2001).  We use a logarithmic radial grid that extends from just outside the inner edge of the disk at $r_{*} = 10^{6}$ cm, out to a radius that safely exceeds the outer edge of the disk at the final time step (typically $\approx 10^{9}$ cm).

The surface density is evolved according to the diffusion equation for an axisymmetric disk in a Newtonian 1/r gravitational potential (e.g., Frank et al.~2002):
\be \frac{\partial\Sigma}{\partial t} = \frac{3}{r}\frac{\partial}{\partial
r}\left[r^{1/2}\frac{\partial}{\partial r}\left(\nu\Sigma r^{1/2}\right)\right],
\label{eq:sigma_evo}
\ee
where the viscosity $\nu$ is proportional to just the gas pressure (see eq.~[\ref{eq:viscosity}] and surrounding discussion).  The radial velocity $v_{r}$ is not evolved explicitly but instead follows from equation (\ref{eq:sigma_evo}) and mass continuity:
\be v_{r} = \frac{-3}{\Sigma r^{1/2}}\frac{\partial}{\partial r}\left(\nu\Sigma r^{1/2}\right).
\ee

We take the initial surface density of the disk to be 
\be \Sigma(r,t=0) \propto \left(\frac{r}{r_{d,0}}\right)^{m}\exp\left[-(2+m)\left(\frac{r}{r_{d,0}}\right)\right], 
\label{eq:sigma0}
\ee
with the constant of proportionality set by requiring that the total disk mass equals $M_{d,0}$.  Equation (\ref{eq:sigma0}) concentrates the initial disk mass $\propto \Sigma r^{2}$ about the radius $r_{d,0}$, with larger values of the parameter $m$ resulting in a more narrowly concentrated mass distribution.  As discussed in $\S\ref{sec:onezoneequations}$, the precise mass distribution of disks produced from CO mergers is uncertain theoretically.  In most of our calculations we take $m = 5$, although our results are relatively insensitive to $m$.  Due to numerical issues that arise from the exponential drop-off of $\Sigma$ at the outer edge of the disk, we impose an initial density floor that is sufficiently small that matter at the density floor contains a fraction $\lesssim 10^{-3}$ of the total disk mass at any time; we have verified that our results are insensitive to the level of this floor as long as it contains negligible mass.  We have also performed calculations using a power-law density distribution that concentrates the disk's initial mass at small radii (e.g., $\Sigma \propto r^{-3}$); we find results that are similar to those obtained using equation (\ref{eq:sigma0}) with $r_{d,0}$ taken near the disk's inner edge.

We calculate the scaleheight $H$ by assuming hydrostatic equilibrium in the vertical direction, i.e., we take $H = c_{s}/\Omega$, where $c_{s} \equiv (P_{\rm tot}/\rho)^{1/2}$ is the isothermal sound speed.  We evolve the midplane temperature using the equation for the specific entropy $S$, which is given by
\be
T\frac{dS}{dt} = \dot{q}_{\rm visc} - \dot{q}_{\nu}^{-} + \dot{q}_{\nu}^{+},
\label{eq:entropy_evo}
\ee
where $(d/dt) \equiv (\partial / \partial t) + v_{r}(\partial / \partial r)$, $\dot{q}_{\rm visc} = (9/4)\nu\Omega^{2}$ is the viscous heating rate, and $\dot{q}_{\nu}^{-}$ is the neutrino cooling rate, as given in Di Matteo et al.~(2002) but modified to include the effects of arbitrary $Y_{e}$ and electron degeneracy (e.g., Beloborodov 2003).  The term 
\bea \dot{q}_{\nu}^{+} = \lambda_{\bar{\nu}_{e}p}\langle\epsilon_{\bar{\nu}_{e}}\rangle\left[Y_{e}-\left(\frac{1-X_{\rm f}}{2}\right)\right]+ \nonumber
\eea
\be
\lambda_{\nu_{e}n}\langle\epsilon_{\nu_{e}}\rangle\left[1-Y_{e}-\left(\frac{1-X_{\rm f}}{2}\right)\right]
\label{eq:nuheating}
 \ee represents neutrino heating from the inner radii of the accretion disk, where $\lambda_{\nu_{e}n}$ and $\lambda_{\bar{\nu}_{e}p}$ are the neutrino capture rates given by equation (\ref{eq:nucaptrate}).  We not not include heating due to $\alpha-$particle formation because we are primarily interested in determining the final electron fraction, and $\alpha-$particles form only after the disk has fallen out of weak equilibrium.  

The initial temperature profile is set by requiring that $H/r = 0.3$ at all radii.  An initially thick disk is physically motivated by the fact that CO merger disks form dynamically hot; however, because the thermal time is short compared to the viscous time, the disk's evolution quickly becomes independent of the initial scaleheight anyways.

Finally, we evolve the electron fraction profile $Y_{e}(r)$ using equation (\ref{eq:yeevo}), with $d/dt = \partial/\partial t + v_{r}\partial/\partial r$.  To keep the required timesteps reasonable, we equate $Y_{e}$ with its equilibrium value $Y_{e}^{eq}$ when the weak interaction rates greatly exceed the local viscous rate 1/$t_{\rm visc}$.  As in $\S\ref{sec:onezonemodel}$, we take the initial electron fraction to be $Y_{e}^{0} = 0.1$ at all radii.

The surface density across the ghost zones at the inner boundary is set to enforce a constant mass accretion rate $\dot{M} \propto \Sigma\nu$ at the value of $\dot{M}$ in the first active zone.  We interpolate the temperature and electron fraction at the inner boundary.  Unlike in the one-zone calculations in $\S\ref{sec:onezonemodel}$, we do not implement a no-torque boundary condition on the inner edge of the disk, in part because it leads to an unphysical temperature profile at small radii (in particular, $T \rightarrow 0$ as $r \rightarrow r_{*}$).  Our results for $Y_{e}^{f}$ are not sensitive to this boundary condition.  We also interpolate all variables at the outer boundary, although this choice has no effect on our results because the outer boundary is not in causal contact with the inner flow.  We have verified that our code conserves total mass $M_{\rm tot}$ and angular momentum $J$ by checking that decreases in $M_{\rm tot}$ and $J$ with time are compensated by their fluxes across the inner grid cell.

\subsection{Results}
\label{sec:COMresults}

\begin{figure}
\begin{center}
\resizebox{\hsize}{!}{\includegraphics[ ]{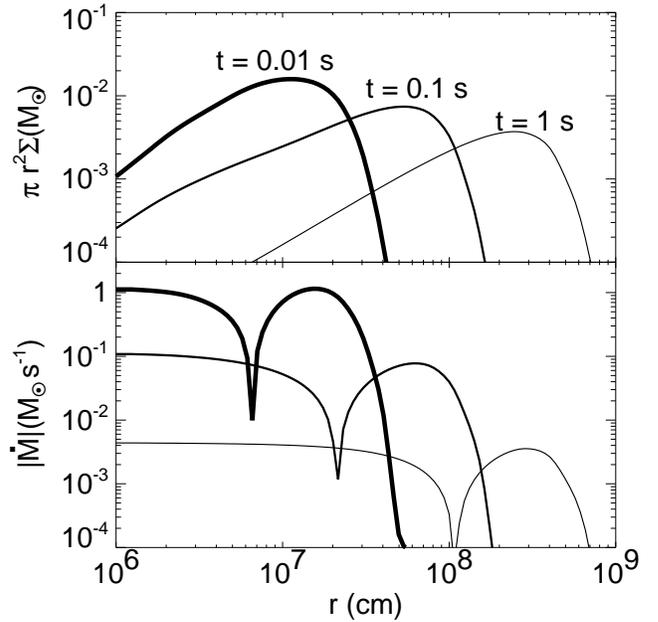}}
\end{center}
\caption{Local disk mass $\pi r^{2}\Sigma$ and the absolute value of the mass advection rate $\dot{M} \equiv 2\pi r v_{r}\Sigma$ as a function of radius.  The model assumes $\alpha = 0.3$, $M_{d,0} = 0.1M_{\sun}$ and $r_{d,0} = 3\times 10^{6}$ cm (with $m=5$ in eq.~[\ref{eq:sigma0}]).  Solutions are shown at $t = 0.01,0.1,$ and 1 s, with later times denoted by increasingly thinner lines.  The nodes in the lower panel occur at the stagnation point that separates matter accreting onto the BH ($\dot{M} < 0$) at small radii from the bulk of the disk that is spreading outwards ($\dot{M} > 0$).  }  
\label{plot:COMdiskevo}
\end{figure}

\begin{figure}
\begin{center}
\resizebox{\hsize}{!}{\includegraphics[ ]{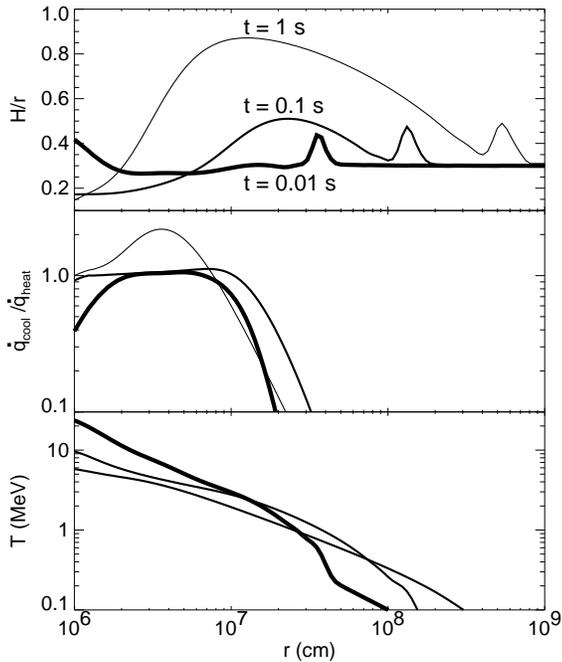}}
\end{center}
\caption{Scaleheight $H/r$, the ratio of neutrino cooling $\dot{q}_{\rm cool} = \dot{q}_{\nu}^{-}$ to total heating $\dot{q}_{\rm heat}$, and the midplane temperature $T$ for the same model and times presented in Figure \ref{plot:COMdiskevo}.  The total heating $\dot{q}_{\rm heat} = \dot{q}_{\rm visc} + \dot{q}_{\nu}^{+}$ consists of viscous heating and neutrino irradiation from the accretion disk at small radii.  }  
\label{plot:COMthermo}
\end{figure}

\begin{figure}
\begin{center}
\resizebox{\hsize}{!}{\includegraphics[ ]{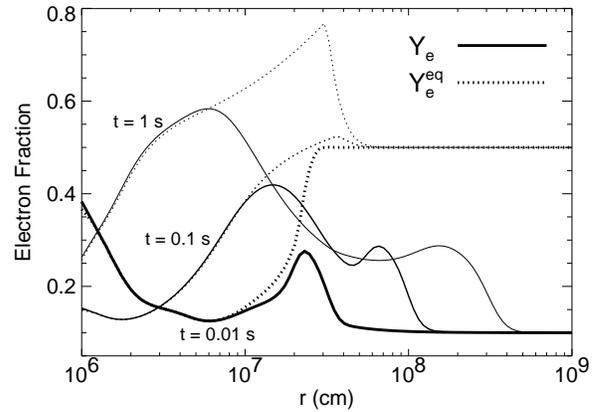}}
\end{center}
\caption{Electron fraction $Y_{e}$ ($\it{solid\,line}$) and the equilibrium electron fraction $Y_{e}^{eq}$ ($\it{dotted\,line}$) for the same model and times presented in Figures \ref{plot:COMdiskevo} and \ref{plot:COMthermo}.  At early times, the disk is in $\beta-$equilibrium, with $Y_{e} \approx Y_{e}^{eq} \ll 0.5$ at all radii that contain substantial mass.  At later times, $Y_{e}^{eq}$ rises as the disk becomes advective and non-degenerate, and $Y_{e}$ freezes out with final values in the range $Y_{e}^{f} \sim 0.1-0.5$ (see Fig.~\ref{plot:2histos}).}
\label{plot:COMye}
\end{figure}

Figures \ref{plot:COMdiskevo}, \ref{plot:COMthermo}, and \ref{plot:COMye} summarize our results for a disk with $\alpha = 0.3$, $M_{d,0} = 0.1M_{\sun}$, and $r_{d,0} = 3\times 10^{6}$ cm, which we take as our fiducial model.  Figure \ref{plot:COMdiskevo} shows the local disk mass $\pi r^{2}\Sigma$ ($\it{top}$) and the absolute value of the mass advection rate $\dot{M} \equiv 2\pi r\Sigma v_{r}$ ($\it{bottom}$) as a function of radius $r$ at times $t = 0.01,0.1,$ and $1$ s, with later times denoted by increasingly thinner lines; these epochs correspond to $\sim$ 1, 10, and 100 times the initial viscous time at $r=r_{d,0}$.

Figure \ref{plot:COMdiskevo} shows that the disk spreads outwards in time, reaching $\sim 3\times 10^{8}$ cm by $t \sim 1$ s.  After a few viscous times, a constant inward accretion rate is established at small radii that roughly matches the {\it outward} mass advection rate of the bulk of the disk.  The inner steady-state disk almost extends to the stagnation point where $v_{r} = 0$, which moves outwards with time.  

Figure \ref{plot:COMthermo} shows radial profiles of the disk thickness $H/r$ ($\it{top}$), the ratio of neutrino cooling $\dot{q}_{\rm cool} = \dot{q}_{\nu}^{-}$ to total heating $\dot{q}_{\rm heat} \equiv \dot{q}_{\rm visc} + \dot{q}^{+}_{\nu}$ (${\it middle}$), and the midplane temperature ($\it{bottom}$).  Figure \ref{plot:COMthermo} illustrates that at early times the bulk of the disk becomes efficiently cooled by neutrinos (i.e., $\dot{q}_{\rm cool} \approx \dot{q}_{\rm heat}$) and geometrically thin ($H/r \ll 1$).  At later times, as the disk spreads and the temperature decreases, neutrinos are no long able to cool the majority of the mass efficiently (i.e., $\dot{q}_{\rm cool} \ll \dot{q}_{\rm heat}$) and $H/r$ increases.  As discussed in MPQ08, the outer disk becomes thick first, and radiatively inefficient conditions move inwards as $\dot{M}$ decreases.  This behavior can be seen explicitly in Figure \ref{plot:COMthermo} by comparing where $H/r$ becomes large at early and late times.  Note that the small ``bump'' in $H/r$ corresponds to low-density material on the very outer edge of the disk which separates the density floor (where the thermal time is always much longer than the evolution timescale) from the bulk of the disk.  This artifact of our initial conditions has no effect on our conclusions.  Also note that at late times ($t = 1$ s) a range of radii around $r \simeq 40$ km has $q_{\rm cool} > q_{\rm heat}$.  This region of net cooling radiates the thermal energy carried into the neutrino-cooled portion of the disk from the higher entropy advective disk at larger radii.

Figure \ref{plot:COMye} shows the evolution of the electron fraction $Y_{e}$ ($\it{solid\,line}$) and its equilibrium value $Y_{e}^{eq}$ ($\it{dotted\,line}$).  At very early times the entire disk is in weak equilibrium, with $Y_{e} = Y_{e}^{eq}$ at all radii that contain substantial mass.\footnote{At radii larger than the extent of the disk, $Y_{e}$ remains equal to its initial value at all times because these regions (which comprise the density floor) never enter equilibrium.}  However, as the disk spreads and thickens, the midplane becomes non-degenerate, which causes $Y_{e}^{eq}$ to rise.  As this occurs, weak interactions become slow compared to the timescale over which the disk evolves (for the same reasons discussed in $\S\ref{sec:onezoneresults}$), and $Y_{e}$ begins to freeze out of equilibrium as it lags behind the rising value of $Y_{e}^{eq}$.  As with radiatively inefficient conditions, freeze-out begins at the outer edge of the disk and moves inwards with time.  By the final time step the majority of the disk mass has frozen out, with values of $Y_{e}^{f}$ that span the range $Y_{e}^{f} \sim 0.1-0.5$.  This behavior is directly analagous to the freeze-out of the ``ring'' radius in the one-zone model (Fig.~\ref{plot:rates_onezone}), but now occuring in each annulus of the disk.

\begin{figure}
\begin{center}
\resizebox{\hsize}{!}{\includegraphics[ ]{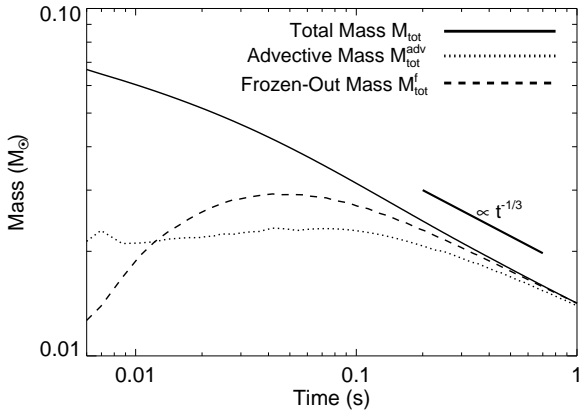}}
\end{center}
\caption{Time evolution of the total mass in the disk $M_{\rm tot}$, the total mass that has become advective because it cannot cool efficiently $M_{\rm tot}^{\rm adv}$ (as defined by eq.~[\ref{eq:advcond}]), and the total mass that has fallen out of weak equilibrium $M^{f}_{\rm tot}$ (as defined by eq.~[\ref{eq:eqcond}]).}  
\label{plot:globalfo}
\end{figure}

In Figure \ref{plot:globalfo} we quantify the global process by which the disk becomes advective and falls out of weak equilibrium by showing the time evolution of the total disk mass $M_{\rm tot}$, the total mass that is advective $M_{\rm tot}^{\rm adv}$, and the total mass that has fallen out of weak equilibrium $M_{\rm tot}^{\rm f}$.  We define annuli that are advective and have fallen out of equilibrium as those that satisfy 
\be \dot{q}_{\rm cool} \lesssim  \dot{q}_{\rm heat}/2 
\label{eq:advcond}
\ee 
and 
\be \lambda_{\rm max} \lesssim 3/t_{\rm visc},
\label{eq:eqcond}
\ee 
respectively, where $\lambda_{\rm max}$ is the maximum of the weak interaction rates in equation (\ref{eq:yeevo}).  Although the numerical prefactors used in equations (\ref{eq:advcond}) and (\ref{eq:eqcond}) are somewhat arbitrary, our results are not sensitive to their precise values.  Figure \ref{plot:globalfo} shows that $M_{\rm tot}$ decreases with time as matter accretes onto the BH, with $M_{\rm tot} \propto t^{-1/3}$ at late times, as expected from the self-similar behavior of an advective disk with $H/r \sim $ constant (MPQ08).  
After an initial viscous time, a comparable amount of material is advective and out of equilibrium; $M_{\rm tot}^{\rm adv}$ and $M_{\rm tot}^{f}$ remain fairly constant at $\sim 2\times 10^{-2}M_{\sun}$ until the majority of the disk becomes advective and freezes out at late times.  

\begin{figure}
\begin{center}
\resizebox{\hsize}{!}{\includegraphics[ ]{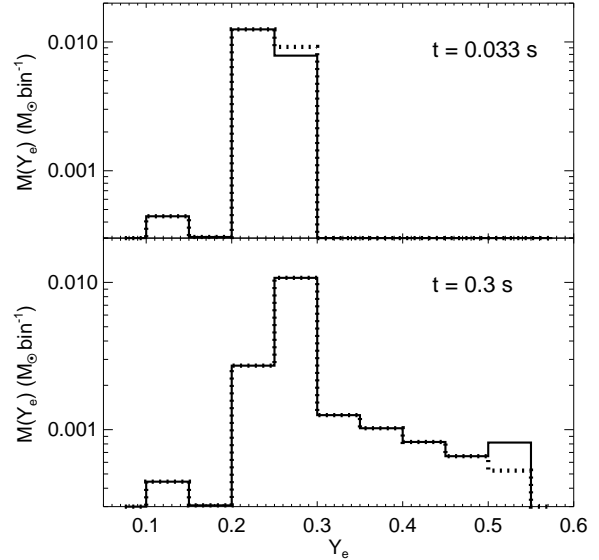}}
\end{center}
\caption{Amount of mass with a given electron fraction $M(Y_{e})$ for the model with $M_{d,0} = 0.1M_{\sun}$, $r_{d,0} = 3\times 10^{6}$ cm, and $\alpha = 0.3$.  The solid and dotted lines include, respectively, mass that has become advective (eq.~[\ref{eq:advcond}]) and that has fallen out of weak equilibrium (eq.~[\ref{eq:eqcond}]).  The times $t = 0.03$ s and 0.3 s correspond to when the disk is $50\%$ and $90\%$ advective, respectively.}  
\label{plot:2histos}
\end{figure}

As discussed in $\S\ref{sec:intro}$, once a portion of the disk becomes advective, a viscously-driven outflow likely drives away a substantial portion of its mass (e.g., Blandford $\&$ Begelman 1999).  In our calculations we have neglected the effects of such a mass sink on the evolution of the disk.  Figure \ref{plot:globalfo} shows that this approximation remains reasonable until $t \sim 0.03-0.1$ s because prior to this point the majority of the disk remains neutrino cooled.  Mass and angular momentum loss to a wind leads to a more rapid decline in $\dot{M}$, which further speeds up the onset of the advective phase and weak freeze-out at smaller radii (MPQ08). 

In Figure \ref{plot:2histos} we show histograms of mass as a function of electron fraction $M(Y_{e})$ at the times $t = 0.03$ s and $t = 0.3$ s.  The solid and dashed lines are the matter that is advective and out of $\beta-$equilibrium, respectively (as defined by eqs.~[\ref{eq:advcond}] and [\ref{eq:eqcond}]).  These distributions typically correspond to the same matter because annuli fall out of equilibrium as they become advective (see eq.~[\ref{eq:foargument}]).  We show the composition at these two particular times because they correspond to epochs when the disk has become $\approx 50\%$ and $90\%$ advective, respectively (Fig.~\ref{plot:globalfo}).  Thus, these distributions likely represent typical values of the electron fraction in the viscously-driven outflows that dominate subsequent mass loss from the disk.

\begin{figure}
\begin{center}
\resizebox{\hsize}{!}{\includegraphics[ ]{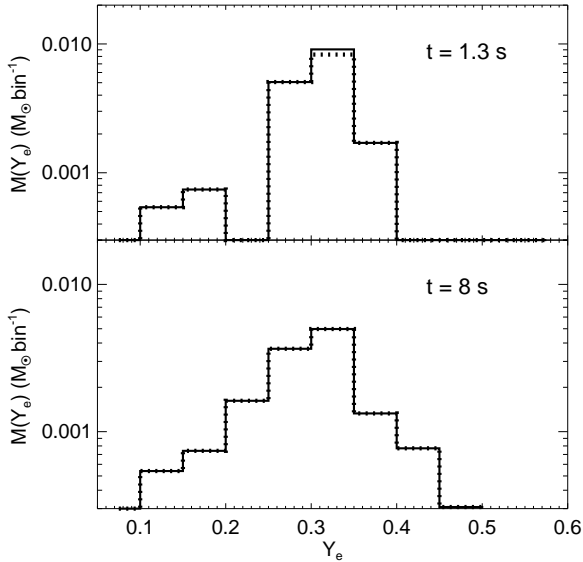}}
\end{center}
\caption{The same as Figure \ref{plot:2histos}, but for $\alpha = 0.03$.  The times shown, $t = 1.3$ s and $8$ s correspond to when the disk is $50\%$ and $90\%$ advective, respectively.}  
\label{plot:2histos_lowalpha}
\end{figure}

It is useful to compare the results in Figure \ref{plot:2histos} with the one zone calculation from $\S\ref{sec:onezonemodel}$ for the same initial disk parameters (see Fig.~\ref{plot:yef_onezone}).  When the disk is $90\%$ advective at $t = 0.3$ s, the height-integrated calculation finds a final electron fraction distribution centered around a mass-averaged value $\langle Y_{e}^{f}\rangle \approx 0.29$, which is somewhat lower than the value of $Y_{e}^{f} = 0.43$ given by the one-zone model.  Indeed, in most of our calculations we find that $Y_{e}^{f}$ calculated using the one zone model tends to slightly exceed $\langle Y_{e}^{f}\rangle$ obtained from the full 1D calculations.

\begin{figure}
\begin{center}
\resizebox{\hsize}{!}{\includegraphics[ ]{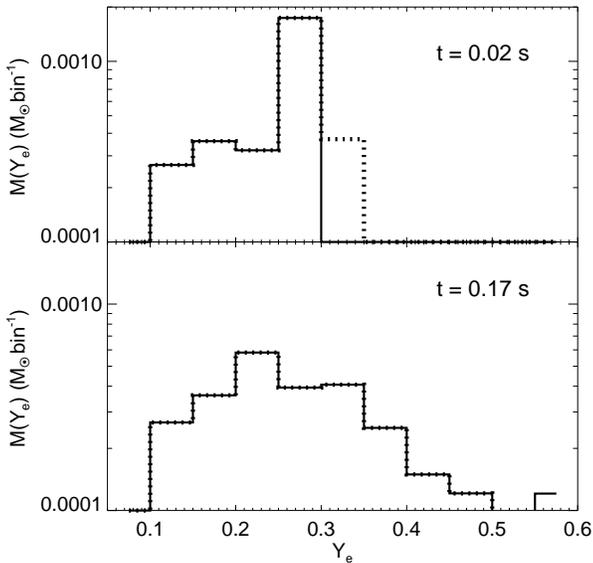}}
\end{center}
\caption{The same as Figure \ref{plot:2histos}, but for $M_{d,0} = 0.01M_{\sun}$.}  
\label{plot:2histos_lowmass}
\end{figure}

In Figures \ref{plot:2histos_lowalpha} and \ref{plot:2histos_lowmass} we show histograms similar to Figure \ref{plot:2histos}, but for solutions calculated with a lower viscosity ($\alpha = 0.03$) and a lower initial mass ($M_{d,0} = 0.01M_{\sun}$), respectively.  The times chosen correspond to epochs when the disk is $50\%$ and $90\%$ advective.  The mass-averaged electron fraction when the disk is $90\%$ advective is $\langle Y_{e}^{f}\rangle \approx 0.3$ for both the lower $\alpha$ and lower $M_{d,0}$ models.  The similarity in the distribution of $Y_{e}^{f}$ for $\alpha = 0.3$ and $\alpha = 0.03$ in Figures $\ref{plot:2histos}$ and $\ref{plot:2histos_lowalpha}$ supports our argument that the freeze-out process is relatively independent of the details of how the disk viscously spreads (see $\S\ref{sec:onezoneresults}$).  The lower disk mass case (Fig.~\ref{plot:2histos_lowmass}) is notable because a significant fraction of the disk's mass freezes out with $Y_{e} \lesssim 0.2$, and such low-$Y_{e}$ material may produce third-peak $r$-process elements (Hoffman et al.~1997).  

We have performed freeze-out calculations for a number of other disk parameters.  In Table \ref{table:ye_avg} we summarize our results for $\langle Y_{e}^{f}\rangle$ and the fraction $f_{\rm adv}$ of the initial disk mass that remains when the disk is $90\%$ advective.  Although different initial mass distributions and viscosities give somewhat different $Y_{e}^{f}$ distributions, the neutron-rich freeze out of $\sim 20-50\%$ of the disk's original mass is a generic property of the disks created during CO mergers.

\section{Discussion}
\label{sec:discussion}

Accretion disks formed from CO mergers that are sufficiently massive and compact to satisfy equation (\ref{eq:md0}) become neutrino-cooled, degenerate, and neutron-rich early in their evolution.  Eventually, as the disk accretes and viscously spreads, neutrino cooling is no longer able to offset viscous heating and the disk becomes advective.  At this point several important changes occur nearly simultaneously: the disk becomes geometrically thick ($H/r \sim 1$), radiation pressure-dominated, and non-degenerate at the same time that the equilibrium electron fraction $Y_{e}^{eq}$ rises from $\ll 0.5$ to $\gtrsim 0.5$ and $Y_{e}$ freezes out (Fig.~\ref{plot:rates_onezone}; see also Beloborodov 2008).  Weak freeze-out necessarily accompanies the disk's advective transition because $e^{-}/e^{+}$ captures are primarily responsible for both cooling the disk and setting its electron fraction (see eq.~[\ref{eq:foargument}] of $\S\ref{sec:onezoneresults}$).  Because $Y_{e}^{eq}$ is rising as freeze-out occurs, moderately neutron-rich freeze-out with $Y_{e}^{f} \approx 0.2-0.4$ is a generic feature of accretion disks formed from CO mergers.  Our calculations in $\S\ref{sec:onezonemodel}$ and $\S\ref{sec:onedmodel}$ largely confirm this basic conclusion (see Table \ref{table:ye_avg}).

\begin{table}
\begin{center}
\vspace{0.05 in}\caption{Properties of Freeze Out in 1D Calculations}
\label{table:ye_avg}

\begin{tabular}{lcccccc}
\hline
\hline
\multicolumn{1}{c}{$M_{d,0}\,(M_{\sun})$} &
\multicolumn{1}{c}{$r_{d,0}\,({\rm cm})$} &
\multicolumn{1}{c}{$\alpha$} &
\multicolumn{1}{c}{$M_{\rm BH}\,(M_{\sun})$} &
\multicolumn{1}{c}{$\langle Y_{e}^{f}\rangle^{(a)}$} &
\multicolumn{1}{c}{$f_{\rm adv}^{(b)}$} &
\\
\hline
\\
0.1 & $3\times 10^{6}$ & 0.3 & 3 & 0.29 & 0.20 \\
- & $3\times 10^{6}$ & 0.03 & 3 & 0.30 & 0.15 \\
- & $6\times 10^{6}$ & 0.3 & 3 & 0.26 & 0.31 \\
- & $1.2\times 10^{7}$ & 0.3 & 3 & 0.21 & 0.40 \\
- & $6\times 10^{6}$ & 0.3 & 10 & 0.37 & 0.28 \\
0.01 & $3\times 10^{6}$ & 0.3 & 3 & 0.28 & 0.28\\
- & $3\times 10^{6}$ & 0.03 & 3 & 0.34 & 0.21 \\
- & $1.2\times 10^{7}$ & 0.3 & 3 & 0.21 & 0.40 \\
0.001 & $3\times 10^{6}$ & 0.3 & 3 & 0.28 & 0.39 \\
- & $1.2\times 10^{7}$ & 0.3 & 3 & 0.17 & 0.72 \\

\hline
\end{tabular}
\end{center}
{\small
(a) Mass-averaged freeze out electron fraction when the disk is $90\%$ advective; (b) Fraction of the initial disk mass that remains when the disk is $90\%$ advective.}
\end{table}

Neutron-rich freeze-out in hyper-accreting disks is usefully contrasted with the ${\it proton}$$-$${\it rich}$ freeze-out ($Y_{e}^{f} \simeq 0.88$) in the very early universe.  Big bang nucleosynthesis (BBN) occurred on a timescale of minutes under non-degenerate and highly radiation-dominated conditions (entropy $S \sim 10^{10}$ $k_{B}$ baryon$^{-1}$).  By contrast, hyper-accreting disks freeze out on a timescale $\sim 0.1-1$ s (depending on $\alpha$) under moderately degenerate conditions, and with comparable gas and radiation pressure, i.e., $S \sim 3-20$ $k_{B}$ baryon$^{-1}$.  Hyper-accreting disks freeze-out neutron-rich because they start in $\beta-$equilibrium under degenerate conditions, while the early universe never had $Y_{e} < 0.5$ because degeneracy effects were never important.
  
A further important difference is in the production of heavy elements.   Once the disk becomes advective and freezes out, a viscously-driven wind likely unbinds most of its remaining mass; as these outflows expand away from the midplane and cool, heavy elements will be formed.  Due to the ``deuterium bottleneck,'' BBN produced almost no elements heavier than He.  By contrast, the late-time outflows from hyper-accreting disks possess modest entropies\footnote{The entropy in the outflow is probably only a few $k_{B}$ baryon$^{-1}$ larger than in the disk midplane because only a fraction of the gravitational binding energy is required to drive an outflow (advective disks are only marginally bound).} and are thus generally in NSE when $\alpha-$particles form.  Even an $\alpha-$rich freeze-out is unlikely (Hoffman $\&$ Woosley 1992) because, under dense (low entropy) neutron-rich conditions, $\alpha$'s burn efficiently via the reaction sequence $^{4}$He($\alpha$ n,$\gamma$)$^{9}$Be($\alpha$,n)$^{12}$C (Delano $\&$ Cameron 1971).  As discussed further in $\S\ref{sec:rates}$, the abundances of the elements synthesized are thus approximately determined by neutron-rich NSE (Hartmann et al.~1985).  

It is also important to contrast the site of nucleosynthesis that we have introduced here with outflows driven from the disk by neutrino heating (e.g., Levinson 2006; Barzilay $\&$ Levison 2008; Metzger et al.~2008b).  Neutrino-driven outflows generally occur at early times or small radii, when the disk midplane is still in $\beta-$equilibrium and the irradiating neutrino flux is substantial.  The outflow's final electron fraction in this case is set by an equilibrium with neutrino absorptions ($Y_{e} \approx Y_{e}^{\nu}$ in eq.~[\ref{eq:yenu}]) and is not directly related to $Y_{e}$ in the midplane (e.g., Surman $\&$ McLaughlin 2004).  By contrast, the outflows considered here occur after the midplane itself has fallen out of weak equilibrium, and so the disk's electron fraction $Y_{e}^{f}$ is retained by the outflow.   While neutrino absorptions in the disk are somewhat important near freeze-out, at no point does the midplane electron fraction enter equilibrium with the neutrinos, as generically occurs in neutrino-driven outflows (Metzger et al.~2008b).  Instead, the outwardly-spreading disk provides a protective ``cocoon'' that transports low-$Y_{e}$ material (kept neutron-rich by degenerate electron capture) away from the deneutronizing flux of neutrinos from the central disk.  Furthermore, the viscously-driven outflows that we have focused on carry away a substantial fraction of the disk's initial mass; because neutrino-driven outflows likely eject less mass (MPQ08), the total nucleosynthetic yield from CO mergers is probably dominated by the viscously-driven outflows at late times.

Finally, although we believe that the disk is largely unbound by viscous heating once it becomes advective (Blandford $\&$ Begelman 1999; Stone $\&$ Pringle 2001), if nothing else the energy released by $\alpha-$particle (or even heavier element) formation is sufficient to unbind the disk at late times (Lee $\&$ Ramirez-Ruiz 2007; MPQ08).  Since $\alpha-$particle formation occurs soon after the disk becomes advective, our conclusions would be qualitatively unchanged if nuclear (rather than viscous) energy powers the late-time outflow.  In this case the formation of heavy elements would begin in the disk midplane, but the final nucleosynthetic yield would probably be similar.

\subsection{Implications}
\label{sec:rates}

As we have argued, moderate entropy, neutron-rich outflows appear to be a common property of accretion disks formed from CO mergers at late times.  Although detailed calculations of the dynamics and nucleosynthesis in these outflows need to be performed, the late-time conditions in the disk favor the production of elements with abundances close to their NSE values at charged-particle reaction freeze-out.  In particular, the entropies  ($\sim 10$ $k_{B}$ baryon$^{-1}$) and dynamical timescales ($\tau_{\rm dyn} \sim 0.01-0.1$ s) at freeze-out are similar to those explored in low entropy nucleosynthesis calculations by Woosley $\&$ Hoffman (1992).  For example, in their model with $S \approx 29$ $k_{B}$ baryon$^{-1}$ and $\tau_{\rm dyn} \approx 0.1$ s, these authors find that all $\alpha-$particles reassemble into heavy nuclei for $Y_{e} \lesssim 0.45$, and that the final abundances closely resemble those of NSE (see the second column in their Table 2).    

Neutron-rich NSE calculations were performed by Hartmann et al.~(1985).  For $Y_{e} \approx 0.35-0.40$, they find that the predominant isotopes created are $^{82}$Se, $^{79}$Br, and $^{78}$Se (in order of decreasing $Y_{e}$; see their Fig.~3).  The very low solar system abundances of these isotopes (total mass fraction $X_{\sun} \approx 3\times 10^{-8}$) strongly constrains the rate of astrophysical events that produce material with $Y_{e} \approx 0.35-0.4$ in NSE: we estimate that the maximum allowed mass ejection of material with $Y_{e} \approx 0.35-0.4$ into the ISM is $\dot{M}_{\rm max} \approx (M_{\rm gal}X_{\sun}/t_{\rm gal}) \approx 1.5\times 10^{-7}M_{\sun}$ yr$^{-1}$, where  $t_{\rm gal} \approx 10^{10}$ yrs and $M_{\rm gal} \approx 5\times 10^{10}M_{\sun}$ and are the age and baryonic mass of the Galaxy, respectively (Binney $\&$ Tremaine 1994).  More neutron-rich material ($Y_{e} \lesssim 0.35$) will produce even heavier isotopes and, possibly, an $r$-process.  Since these elements are also very rare, a similar constraint can be placed on the rate that very low-$Y_{e}$ material is ejected into the ISM.  For instance, isotopes near the second $r$-process peak ($A \sim 130$) have a typical solar mass abundance $X_{\sun}^{r} \sim 10^{-9}$ relative to hydrogen; therefore, if material ejected with $0.2 \lesssim Y_{e} \lesssim 0.35$ produced equal numbers of $N \sim 10$ $r$-process isotopes in this mass range, the constraint on the mass ejected would be $\dot{M}_{\rm max} \sim 5\times 10^{-8}(N/10)M_{\sun}$ yr$^{-1}$, comparable to the rate for material with $Y_{e} \approx 0.35-0.4$ given above.  In order to be conservative, below we use the constraint $\dot{M}_{\rm max} \approx 1.5\times 10^{-7}M_{\sun}$ yr$^{-1}$ for all material ejected with $Y_{e} \lesssim 0.4$.  However, a more precise determination of the isotopes produced by the ejection of moderate entropy material with $Y_{e} \approx 0.2-0.35$ would likely place even stronger constraints on the rate of CO mergers.

Based on Figures \ref{plot:2histos}-\ref{plot:2histos_lowmass} and our one-zone calculations in $\S\ref{sec:onezonemodel}$ we find that a fraction $f_{\rm adv} \approx 0.2-0.5$ of the disk's initial mass remains when it becomes advective at late times (see Table \ref{table:ye_avg}), with a higher fraction for less massive, more extended disks.  Of this material, we estimate that the majority of the mass will have $Y_{e}^{f} \lesssim 0.4$, when averaging over many events with different initial disk properties (see Figs.~\ref{plot:2histos}-\ref{plot:2histos_lowmass} and Table \ref{table:ye_avg}).  Although ejecta from mergers that occur outside Galactic disk will not enrich the ISM, population synthesis studies estimate that for large spirals such as the Milky Way only a small fraction ($\lesssim 20\%$) of mergers occur in intergalactic space (Belczynski et al.~2006).  Thus a fraction $\eta \approx 0.1-0.5$ of the disk's initial mass will both become unbound from the disk and pollute the ISM with rare neutron-rich isotopes.  Hence, if the average mass of the disks formed from CO mergers is $\langle M_{d,0} \rangle$, their rate in the Milky Way cannot exceed
\be
\dot{N}_{\rm max} \approx \frac{\dot{M}_{\rm max}}{\eta \langle M_{d,0}\rangle} \approx 10^{-5}\left(\frac{\eta}{0.2}\right)^{-1}\left(\frac{\langle M_{d,0} \rangle}{0.1M_{\sun}}\right)^{-1} {\rm yr^{-1}},
\label{eq:maxrate}
\ee
where we have normalized $\eta$ to a typical value $\sim 0.2$.

Based on observed binary NS systems, Kalogera et al.~(2004) find that the NS-NS merger rate in the Milky Way is between $1.7\times 10^{-5}$ and $2.9\times 10^{-4}$ yr$^{-1}$ at 95$\%$ confidence.  Population synthesis estimates (e.g., Belczynski et al.~2006) are consistent with this range but with larger uncertainties.  Our constraint in equation (\ref{eq:maxrate}) thus requires either NS-NS merger rates at the low end of current estimates, or a small average disk mass $\ll 0.1M_{\sun}$.  

Ultimately, equation (\ref{eq:maxrate}) must be combined with GR simulations of the merger process that determine $M_{d,0}$ as a function of the total binary mass $M_{\rm tot} = M_{1} + M_{2}$, the NS mass ratio $q = M_{1}/M_{2} \leq 1$, and the NS equation of state (EOS).  General relativistic merger simulations (Shibata et al.~2005; Shibata $\&$ Taniguchi 2006) find that when $M_{\rm tot}$ is above a critical value ($\gtrsim 2.6M_{\sun}$, depending on the EOS) the central object produced during the merger promptly collapses to a BH.  In this case, $M_{d,0}$ decreases rapidly with increasing $q$ (see Shibata $\&$ Taniguchi 2006, Fig.~13).  For instance, when the double pulsar system PSR J0737-3039A/B (with $q \simeq 0.935$ and $M_{\rm tot} = 2.587M_{\sun}$) merges in $\approx 100$ Myr, current simulations predict that the disk will have a mass $\sim 10^{-3}-10^{-2}M_{\sun}$ if BH formation is prompt (Shibata $\&$ Taniguchi 2006).  This disk mass is reasonably consistent with our constraint in equation (\ref{eq:maxrate}) and current NS-NS merger rate estimates.  It is also consistent with the relatively low isotropic energies of short GRBs with measured redshifts ($\sim 10^{-5}-10^{-3}M_{\sun}$ c$^{2}$; Nakar 2007).  

On the other hand, simulations find that when $M_{\rm tot}$ is below the threshold for prompt collapse, a hypermassive NS supported by differential rotation is initially formed (Baumgarte et al.~2000).  If this NS is able to transport angular momentum outwards, it will collapse to a BH on a longer timescale ($\approx 100$ ms; Shibata et al.~2006; Duez et al.~2006), producing a sizable disk ($\gtrsim 0.03M_{\sun}$) whose mass is relatively independent of $q$ (Oechslin $\&$ Janka 2006; Shibata et al.~2006).  Since the disk masses formed after delayed collapse are comparable to those allowed by our constraint in equation (\ref{eq:maxrate}) even for low $\dot{N} \sim 10^{-5}$ yr$^{-1}$, our results suggest that the formation of a long-lived hyper-massive NS is a rare outcome of NS-NS mergers.

Since there are no known BH-NS binaries, the BH-NS merger rate is even less certain.  Bethe $\&$ Brown (1998) argue that BH-NS mergers could be substantially more common than NS-NS mergers, with Bethe et al.~(2007) estimating a rate $\sim 10^{4}$ Gpc$^{-3}$yr$^{-1}$, corresponding to $\sim 10^{-3}$ yr$^{-1}$ in the Milky Way.  Our results rule out such a high rate unless $\langle M_{d,0} \rangle$ is less than $\sim 10^{-3}M_{\sun}$.  A low average disk mass may be possible, however, because only systems for which the NS-BH mass ratio falls within a fairly narrow range may produce a sizable disk upon merging (e.g., Miller 2005). 

The late-time nucleosynthesis in CO merger disks also places interesting constraints on the beaming fraction of short GRBs if most of these events truly result from CO mergers.  Using the observed local short GRB rate of $\approx 10$ Gpc$^{3}$ yr$^{-1}$ (Nakar et al.~2006) and assuming that the merger rate is proportional to the blue stellar luminosity (Phinney 1991), the short GRB rate in the Milky Way is estimated to be $\dot{N}_{\rm SGRB} \approx 10^{-6}$ yr$^{-1}$ (Nakar 2007).  Thus, if all mergers that produce disks also produce short GRBs, the beaming fraction must obey
\be
f_{b} \gtrsim \frac{\dot{N}_{\rm SGRB}}{\dot{N}_{\rm max}} \approx 0.13 \left(\frac{\eta}{0.2}\right)\left(\frac{\langle M_{d,0}\rangle }{0.1M_{\sun}}\right).
\label{eq:minbeaming}
\ee
This constraint implies that the jet half-opening angle must exceed $\theta \sim (2f_{b})^{1/2} \sim 30^{\circ}(\eta/0.2)^{1/2}(M_{d,0}/0.1M_{\sun})^{1/2}$.  This is consistent with other evidence that short GRBs may be less collimated than long GRBs (e.g., Grupe et al.~2006; Soderberg et al.~2006), which are instead inferred to have $0.002 \lesssim f_{b} \lesssim 0.01$ (Frail et al.~2001; Bloom et al.~2003; Guetta, Piran, $\&$ Waxman 2005).  Note that equation (\ref{eq:minbeaming}) is probably a stronger constraint than equation (\ref{eq:maxrate}) because only CO mergers that form sizable accretion disks are likely to produce GRBs in the first place. 

In addition to material driven from the disk by late time winds, neutron-rich matter may be ejected ${\it dynamically}$ during the merger process itself (e.g., Freiburghaus et al.~1999; Rosswog et al.~1999).  Although the amount of material ejected in merger calculations depends sensitively on details such as the NS spin and equation of state (Rosswog et al.~1999), the rare isotopes synthesized by this highly neutron-rich ejecta may provide a constraint on the CO merger rate that is comparable to that from late-time outflows given by equation (\ref{eq:maxrate}).

Our conclusion that accretion disks formed from CO mergers freeze out neutron-rich is also consistent with the absence of SN-like longer wavelength transients coincident with these events.  If CO merger disks froze out with $Y_{e}^{f} \gtrsim 0.5$ instead of $Y_{e}^{f} \lesssim 0.5$, their late-time outflows would efficiently synthesize $^{56}$Ni (e.g., Seitenzahl et al.~2008), which could power an optical/infrared transient on $\sim$ day timescales following short-duration GRBs.   Despite intensive searches, however, such transients have not yet been observed.  For instance, early optical follow-up of the GRB 050509b (Hjorth et al.~2005) limited the amount of $^{56}$Ni to $\lesssim 10^{-2}M_{\sun}$ (Kulkarni 2005; MPQ08), consistent with the modest amount of material with $Y_{e}^{f} \gtrsim 0.5$ that we find in Figures \ref{plot:2histos}-\ref{plot:2histos_lowmass}.  Although we do not expect much $^{56}$Ni to be produced in late-time outflows from CO merger disks, earlier neutrino-driven outflows may produce up to $\sim 10^{-3} M_{\sun}$ in Ni, which could power a fainter optical or infrared transient (MPQ08).  In addition, outflows powered by fall-back accretion onto the BH at late times could power an X-ray transient on a timescale of days to weeks (Rossi $\&$ Begelman 2008).
\section*{Acknowledgments}

We thank Todd Thompson and Stan Woosley for helpful conversations and useful information.  A.L.P. is supported by the Theoretical Astrophysics Center at UC Berkeley.  B.D.M.  and E.Q. are supported in part by the David and Lucile Packard Foundation and a NASA GSRP Fellowship to B.D.M.

\label{lastpage}

\end{document}